\documentstyle[aps]{revtex}
\begin{document}
\title{The elastic QCD dipole amplitude at one-loop}
\author{H. Navelet and R. Peschanski\thanks{%
CEA, Service de Physique Theorique, CE-Saclay, F-91191 Gif-sur-Yvette Cedex,
France}}
\maketitle

\begin{abstract}
We derive the analytic expression of the two one-loop dipole contributions
to the elastic 4-gluon amplitude in QCD. The first one corresponds to the
double QCD pomeron exchange, the other to an order $\alpha^2$ correction to
one-pomeron exchange. Both are expressed in terms of the square of the
recently derived triple QCD pomeron vertex and involve a summation over all
conformal Eigenvectors of the BFKL kernel.
\end{abstract}
\bigskip
{\bf 1.} The bare (BFKL) pomeron in Quantum chromodynamics has been derived
long ago \cite{bfkl} by resummation of the leading $\left(\bar \alpha \log
(1/x) \right)^n$ contributions to the QCD perturbative expansion at small $x,
$ $\bar \alpha$ being an arbitrarily fixed but small QCD coupling. This
derivation allows to compute the 4-gluon elastic amplitude ${\cal A}_Q
(k,k^{\prime};Y),$ where $k,k^{\prime}$ are the two-dimensional transverse
components of the initial momenta, $Q,$ the momentum transferred in the
elastic reaction, and the total available rapidity interval $Y (\equiv \log
(1/x) {\rm \ in\ DIS\ amplitudes)}.$ It is well-known that the bare BFKL
pomeron has an energy dependence violating the Froissart bound, which is a
basic consequence of unitarity and the existence of a minimum pion mass. It
has been suggested that the unitarity constraint could be satisfied within
the weak coupling regime of QCD through the computation of multi-pomeron
exchange \cite{mu1}. A first approximate attempt of calculating the double
pomeron exchange has been performed \cite{mu3}. Numerical estimates of the
multi-pomeron contributions have also been presented \cite{salam} . However,
the corresponding exact analytic expressions were not yet available.

The aim of our paper is to give the analytical expression of the one-loop
dipole contribution to the elastic amplitude in the QCD-dipole picture of
BFKL dynamics \cite{mu2}. The QCD dipole formulation is known \cite{equiv}
to be equivalent at tree level to the $SL(2,{\cal C})$-invariant BFKL
amplitude in terms of Feynman graphs. The idea is to use the dipole
formalism as an effective theory defining the propagation and interaction
vertices of two QCD Pomerons, which are compound states of reggeized gluons
in the BFKL \cite{bfkl} representation.

A first attempt \cite{mu3} has been made to compute the double-Pomeron
contribution to the forward elastic amplitude as a first-order unitarity
correction to the onium-onium total cross-section at large $Y.$ However,
this approach was using approximate expressions valid at large
impact-parameter only, and thus the whole content of $SL(2,{\cal C})$
symmetry was lost. In the present paper we derive the exact expressions,
without approximation, and valid for any $Q$ (thus for both forward and
non-forward amplitudes).

Let us first introduce the $SL(2,{\cal C})$-invariant formalism for the
4-gluon elastic amplitude ${\cal A}_Q (k,k^{\prime};Y)$ in the BFKL
derivation. The solution of the BFKL equation is more easily expressed in
terms of the Fourier transformed amplitude $f_Q (\rho,\rho^{\prime};Y)$
given by the relation 
\begin{equation}
{\cal A}_Q (k,k^{\prime};Y)=\frac 1{(2\pi)^5}\int d^2\!\rho
d^2\!\rho^{\prime}\ e^ {i\rho\left(k-\frac Q2 \right)-i\rho^{\prime}%
\left(k^{\prime}-\frac Q2 \right)} f_Q (\rho,\rho^{\prime};Y).  \label{2}
\end{equation}
Using the $SL(2,{\cal C})$-invariant formalism, the solution of the BFKL
equation reads \cite{lip} 
\begin{equation}
f_Q (\rho,\rho^{\prime};Y)=\bar \alpha ^2\ \frac {\mid\rho\rho^{\prime}\mid}{%
16} \int dh \ \bar E^h_Q (\rho^{\prime}) E^h_Q (\rho)\ {d(h)}e^{\omega(h)Y},
\label{4}
\end{equation}
where the factor $\bar \alpha ^2$ comes from the coupling to incident
dipoles. In equation (\ref{4}), the symbolic notation $\int dh \equiv
\sum^{\infty}_{n=-\infty} \ \int d\nu $ corresponds to the integration over
the $SL(2,{\cal C})$ quantum numbers with $h=i\nu + \frac {1-n}2.$ $E^h_Q
(\rho)$ and $\omega(h)$ are, respectively, the $SL(2,{\cal C})$
Eigenfunctions and Eigenvalues of the BFKL kernel \cite{lip}. The
Eigenvalues read 
\begin{equation}
\omega(h) = \frac {\bar \alpha N_c}{\pi} \ \chi (h) \equiv \frac {\bar \alpha
N_c}{\pi}\ 2\left\{\Psi(1)-\Re \left(\Psi\left(\frac {1\!+\!\mid\!n\!\mid}2
+i\nu\right)\right)\right\},  \label{6}
\end{equation}
where $\Psi \equiv (\log\Gamma)^{\prime}.$ The $SL(2,{\cal C})$ Eigenvectors
are defined by 
\begin{equation}
E^h_Q (\rho) = \frac {2\pi^2}{\mid\!\rho\!\mid b(h)}\int d^2b\ e^{iQ\cdot
b}\ E^h \left(b\!-\!\frac{\rho}2,b\!+\!\frac{\rho}2\right),  \label{8}
\end{equation}
with 
\begin{equation}
E^h \left(b\!-\!\frac{\rho}2,b\!+\!\frac {\rho}2 \right)=(-)^{h-\tilde h}
\left(\frac{\rho}{b^2\!-\!\frac{\rho ^2}4}\right)^h\ \left(\frac{\bar \rho}{%
\bar b^2\!-\!\frac{\bar \rho ^2}4}\right)^{\tilde h},  \label{10}
\end{equation}
where $\tilde h = 1-\bar h,$ $b$ is the 2-d impact-parameter, and 
\begin{equation}
d(h) = \left\{\left[\nu ^2 \!+\! \frac {(n\!-\!1)^2}4\right] \left[\nu ^2
\!+\! \frac {(n\!+\!1)^2}4\right]\right\}^{-1},\ b(h)=\frac {\pi^3 4^{h+%
\tilde h -1}}{\frac 12 -h}\ \frac {\gamma(1\!-\!h)}{\gamma(\frac 12\!-\!h)},
\label{12}
\end{equation}
with $\gamma (z) \equiv \frac {\Gamma (z)}{\Gamma (1-\tilde z)}.$ Note that
an analytic expression of the Eigenvectors $E^h_Q (\rho)$ in the mixed
representation has been provided \cite{equiv} in terms of a combination of
products of two Bessel functions. For simplicity, we did not include the
impact factors corresponding to the external coupling (see \cite{lip,vertex}%
). Note also that the leading contribution of the amplitude (\ref{4}) is the 
$n\!=\!0$ component which corresponds to the BFKL QCD Pomeron.

The final expressions to be  obtained in this letter read: 
\begin{eqnarray}
f^{({\cal P})}_Q (\rho,\rho^{\prime};Y) &=& \bar \alpha ^4 \int dh \ E^h_Q
(\rho^{\prime}) \bar E^h_Q(\rho)\int dh_a dh_b\ \left\vert \frac {g_{3{\cal P%
}} (h,h_a,h_b)} {b(h_a)b(h_b)b(h)} \right\vert^2  \nonumber \\
& &\frac {e^{\omega(h) Y}}{(\chi (h)-\chi (h_a)-\chi (h_b))^2 }\ ,
\label{14}
\end{eqnarray}
\begin{eqnarray}
f^{({\cal P}\otimes{\cal P})}_Q (\rho,\rho^{\prime};Y) &=& \bar \alpha ^4
\int dh \ E^h_Q (\rho^{\prime}) \bar E^h_Q(\rho)\int dh_a dh_b\ \left\vert 
\frac {g_{3{\cal P}} (h,h_a,h_b)} {b(h_a)b(h_b)b(h)} \right\vert^2  \nonumber
\\
&\times&\frac {e^{\left(\omega(h_a)+\omega(h_b)\right) Y}}{(\chi (h)-\chi
(h_a)-\chi (h_b))^2 }\ ,  \label{16}
\end{eqnarray}
where $g_{3{\cal P}}$ is the triple-QCD-Pomeron vertex (e.g. $1\to 2$ dipole vertex
in the $1/N_C$ limit) which has been recently derived \cite{triplep}.

The two contributions correspond respectively to the one-dipole-loop
correction to the BFKL Pomeron ($f^{({\cal P})}_Q$) and to the two-Pomeron
exchange ($f^{({\cal P}\otimes{\cal P})}_Q$). Indeed, the energy dependence
of these contributions is fixed by the asymptotic behaviour in $Y$ of
formulae (\ref{14},\ref{16}), which we will show later to be, respectively,
in the vicinity of the one-Pomeron and two-Pomeron intercepts. Notice that
the forward amplitudes ($Q=0$) are obtained by replacing $E^h_Q(\rho) \to
(\rho)^{\frac 12-h}.$ Formulae (\ref{14},\ref{16}) are the main results of
this paper. We will comment these results and compare with the existing
approximate evaluations in the further discussion.

\bigskip
{\bf 2.} Let us come to the derivation of formulae (\ref{14},\ref{16}). The
formulation of the general one-loop amplitude in the QCD dipole model can be
written \cite {mu3} : 
\begin{eqnarray}
& &f^{(one-loop)} \left( \rho _{0}\rho
_{1};\rho^{\prime}_{0}\rho^{\prime}_{1} | Y\!=\!y\!+\!y^{\prime}\right)= 
\frac 1{2!(2\pi)^8}  \nonumber \\
& & \int_0^y d \bar y \int_0^{y^{\prime}} d \bar y^{\prime}\int \frac{%
d^{2}\rho _{a_0} d^{2}\rho _{a_1}d^{2}\rho _{b_0}d^{2}\rho _{b_1}}{\left|
\rho _{a}\ \rho _{b}\right| ^{2}} \frac{d^{2}\rho _{a^{\prime}_0}d^{2}\rho
_{a^{\prime}_1}d^{2}\rho _{b^{\prime}_0}d^{2}\rho _{b^{\prime}_1}}{\left|
\rho _{a^{\prime}}\ \rho _{b^{\prime}}\right| ^{2}}  \nonumber \\
& &n_{2}\left( \rho _{0}\rho _{1};\rho _{a_0}\rho _{a_1},\rho _{b_0}\rho
_{b_1} | y\!-\!\bar y,\bar y\right) n_{2}\left(
\rho^{\prime}_{0}\rho^{\prime}_{1};\rho^{\prime}_{a_0}\rho^{\prime}_{a_1},%
\rho^{\prime}_{b_0}\rho^{\prime}_{b_1} | y^{\prime}\!-\!\bar y^{\prime},\bar %
y^{\prime}\right)  \nonumber \\
& & \ \ \ \ \ \ \ T(\rho _{a_0}\rho _{a_1},\rho _{a^{\prime}_0}\rho
_{a^{\prime}_1})\ T(\rho _{b_0}\rho _{b_1},\rho _{b^{\prime}_0}\rho
_{b^{\prime}_1}),  \label{18}
\end{eqnarray}
where $\rho _{0}\rho _{1}$ are the transverse coordinates of one of the
initially colliding dipoles (resp. $\rho^{\prime}_{0}\rho^{\prime}_{1}$ for
the second one), $\rho _{a_0}\rho _{a_1}$ and $\rho _{b_0}\rho _{b_1},$ the
two interacting dipoles emerging from the dipole $\rho _{0}\rho _{1}$ after
evolution in rapidity (resp. $\rho_{i} \to \rho_{i^{\prime}}$, for the
second one), see figure {\bf 1a,b}. It is important to notice that one has
to introduce the probability distributions $n_{2}(| y\!-\!\bar y,\bar y)$ of
producing two dipoles after a {\it mixed} rapidity evolution, namely with a
rapidity $y\!-\!\bar y$ with one-Pomeron type of evolution and a rapidity $%
\bar y$ with two-Pomeron type of evolution and integrate over $\bar y.$ The
interaction amplitudes $T(\rho _{a_0}\rho _{a_1},\rho _{a^{\prime}_0}\rho
_{a^{\prime}_1})$ and $T(\rho _{b_0}\rho _{b_1},\rho _{b^{\prime}_0}\rho
_{b^{\prime}_1})$ are the elementary two-gluon exchange amplitudes between
two colorless dipoles, namely 
\begin{equation}
T(\rho _{a_0}\rho_{a_1},\rho _{a^{\prime}_0}\rho_{a^{\prime}_1})= \int d^2q
\ e^{i\frac q2 \left(\rho _{a_0}\!+\!\rho_{a_1}\!-\!\rho
_{a^{\prime}_0}\!-\!\rho_{a^{\prime}_1}\right)} \ f_q (\rho
_{a_0}\!-\!\rho_{a_1},\rho _{a^{\prime}_0}\!-\!\rho_{a^{\prime}_1};Y\!=\!0).
\label{19}
\end{equation}
$n_2$ obeys a mixed evolution equation \cite{mu3,n2} the solution of
which is a mere extension of the one formulated in ref. \cite{pe}. It reads 
\begin{eqnarray}
n_{2}\left( \left. \rho _{0}\rho _{1};\rho _{a_0}\rho _{a_1},\rho _{b_0}\rho
_{b_1}\right|y-\bar y,\bar y\right) =\frac {\bar \alpha N_C}{\pi}\ \int 
\frac {dh dh_a dh_b}{\mid \rho_{}\rho_a\rho_b\mid^2}  \nonumber \\
\int d\omega_1 \frac {e^{\omega_1 (y-\bar y)}} {2a(h)\left( \omega_1-\omega
\left( h\right) \right)} \int d\omega \frac {e^{\omega y}} {a(h_{a})
a(h_{b})\ \left(\omega \left( h_{a}\right) +\omega \left( h_{b}\right)
-\omega \right) }  \nonumber \\
\times \int d^{2}\rho _\alpha d^{2}\rho _\beta \ d^{2}\rho _\gamma {E}%
^{h_{a}}{\left( \rho _{a_0 \alpha} ,\rho _{a_1\alpha} \right) }\ {E}^{h_{b}}{%
\left( \rho _{b_0\beta} ,\rho _{b_1\beta} \right) } {E}^{h}{\left( \rho _{0
\gamma} ,\rho _{1\gamma} \right) } {\bar {{\cal R}}}^{h,h_a,h_b}_{\alpha,%
\beta,\gamma},  \label{20}
\end{eqnarray}
with 
\begin{equation}
{\cal R}^{h,h_a,h_b}_{\alpha,\beta,\gamma}\equiv \int \frac {%
d^{2}r_{0}d^{2}r_{1}d^{2}r_{2}}{\left| r_{01}\ r_{02}\ r_{12}\right| ^{2}} \
E^{h}{\left( r_{0\gamma},r_{1\gamma}\right)}E^{h_{a}}{\left( r_{0\alpha
},r_{2\alpha }\right) }E^{h_{b}}{\left( r_{1\beta} ,r_{2\beta} \right) }
\label{22}
\end{equation}
where $\rho =\rho _{0}\!-\!\rho _{1},\rho _{a}=\rho _{a_0}\!-\!\rho _{a_1},\
\rho _{b}=\rho _{b_0}\!-\!\rho _{b_1}.$

Conformal $SL(2,{\cal C})$ invariance implies 
\begin{eqnarray}
{\cal R}^{h,h_a,h_b}_{\alpha,\beta,\gamma}\equiv
\left[\rho_{\alpha\beta}\right]^{h-h_a-h_b}
\left[\rho_{\beta\gamma}\right]^{h_a-h_b-h}
\left[\rho_{\gamma\alpha}\right]^{h_b-h_a-h}  \nonumber \\
\left[\bar\rho_{\alpha\beta}\right]^{\tilde h-\tilde h_a-\tilde h_b} \left[%
\bar \rho_{\beta\gamma}\right]^{\tilde h_a-\tilde h_b-\tilde h} \left[\bar%
\rho_{\gamma\alpha}\right]^{\tilde h_b-\tilde h_a-\tilde h}\ g_{3{\cal P}%
}\left(h,h_a,h_b\right),  \label{24}
\end{eqnarray}
where $g_{3{\cal P}}\left(h,h_a,h_b\right)$ is the triple Pomeron coupling
in the $1/N_C$ limit  obtained in the QCD dipole model, namely: 
\begin{eqnarray}
g_{3{\cal P}}(h,h_a,h_b)= \int \frac {d^{2}r_{0}d^{2}r_{1}d^{2}r_{2}}{\left|
r_{01}\ r_{02}\ r_{12}\right| ^{2}} \left[r_{01}\right]^{h} \left[\frac {%
r_{02}}{r_0r_2}\right]^{h_a}\ \times  \nonumber \\
\times \ \left[\frac {r_{12}}{\left(1-r_1\right)\left(1-r_2\right)}%
\right]^{h_b}\left[\bar r_{01}\right]^{\tilde h} \left[\frac {\bar r_{02}}{%
\bar r_0\bar r_2}\right]^{\tilde h_a} \left[\frac {\bar r_{12}}{\left(1-\bar %
r_1\right) \left(1-\bar r_2\right)}\right]^{\tilde h_b}.  \label{26}
\end{eqnarray}
From general arguments of conformal invariance, it is enough to calculate
the forward elastic amplitude ($Q=0$), since the $Q$-dependence is given
exclusively by the expansion over the conformal eigenvectors $\int dh \
E^h_Q (\rho^{\prime})\ \bar E^h_Q(\rho).$ This property can be directly
checked by a tedious but explicit calculation, not reproduced here (see \cite
{nav2}).

It is convenient to introduce the double Fourier transform of the $1\to 2$
dipole distribution. One defines 
\begin{equation}
\tilde n_2(\rho,\!\rho_a, \!\rho_b;\!q_a, \!q_b\vert y-\!\bar y,\!\bar y%
)\equiv \! \int \! e^{iq_a \!b_a+q_b\! b_b}d^2b_a d^2b_b \ n_{2}\left(\!\rho
_{0}\rho _{1};\!\rho _{a_0}\rho _{a_1}\!,\!\rho _{b_0}\rho
_{b_1}\!\mid\!y-\!\bar y,\!\bar y\right),  \label{28}
\end{equation}
where $q_a$ (resp.$q_b$) is the tranverse momentum of the dipole $%
\rho_{a0}\rho_{a1}$ (resp. $\rho_{b0}\rho_{b1}$) with respect the forward
direction and $2 b_a = \rho _{a_0}+\rho _{a_1},$ (resp. $2 b_b = \rho
_{b_0}+\rho _{b_1}$), is the impact parameter.

Indeed, the computation of the forward amplitude ($Q= q_a \!+\!q_b \!=\! 0$)
requires only the simpler solution for $\tilde n_2(\rho,\rho_a,
\rho_b;q, -q\vert y\!-\!\bar y,\bar y).$ Inserting formula (\ref{20}) in
definition (\ref{28}) the integration over $d^{2}\rho _\alpha d^{2}\rho
_\beta \ d^{2}\rho _\gamma $ can be performed using the Eigenvectors (\ref{8}%
) in the mixed representation. After removing the $\delta$-function of
transverse momentum conservation, we get: 
\begin{eqnarray}
&& \tilde n_2(\rho,\rho_a, \rho_b;q, -q\vert y\!-\!\bar y,\bar y)= 
\frac {\alpha N_C}{\pi}\ \int \frac {dh dh_a dh_b}{\mid \rho_{
}\rho_a\rho_b\mid^2} \ {\cal G}(h,h_a,h_b)  \nonumber \\
&\times& \int d\omega_1 \frac {e^{\omega_1 (y-\bar y)}}{\left(
\omega_1-\omega \left( h\right) \right)} \int d\omega \frac {e^{\omega y}} {%
\left(\omega \left( h_{a}\right) +\omega \left( h_{b}\right) -\omega \right) 
}  \nonumber \\
&\times& \left[\rho\right]^{\frac 12 -h} \left[\bar \rho\right]^{\frac 12 -%
\tilde h} \left(\frac q2\right)^{1-\tilde h_a-\tilde h_b-\tilde h} \left(%
\frac {\bar q}2\right)^{1-h_a-h_b-h} {E}^{h_{a}}_q \left( \rho_a\right) {E}%
^{h_{b}}_{-q} \left( \rho_b\right) ,  \label{30}
\end{eqnarray}
where 
\begin{equation}
{\cal G}(h,h_a,h_b) = \frac {\bar g_{3{\cal P}}(h,h_a,h_b)}{\bar b(h_a)\bar b%
(h_b)\bar b(h)}\frac {\gamma(h_a\!+\!h_b\!+\!h\!-\!1)\gamma(h_a\!-\!h_b\!+%
\!h)\gamma(h_b\!-\!h_a\!+\!h)}{\gamma(2h)},  \label{32}
\end{equation}
and, by definition, $\gamma (z) \equiv \frac {\Gamma (z)}{\Gamma (1-\tilde z)%
}\ .$ To perform the eight integrals over transverse coordinates in the
formula (\ref{18}) for the one-loop amplitude we use the expression (\ref{20}%
) for $n_2$ and  introduce the double Fourier transform (cf. $\tilde n%
_2(q_a,q_b),$ see (\ref{28})) and the one-dimensional Fourier transform
of the interaction amplitudes $T,$ see (\ref{19}). The integral over the
intermediate impact parameters leads to $\delta$-functions of the various
transverse momenta ($q_a,q_{a^{\prime}},$ etc...) while the remaining
integrations over the differences $\rho_i-\rho_j$ give rise to $\delta$%
-functions over the conformal weights, according to the known orthogonality
properties \cite{lip,equiv} of the $SL(2,{\cal C})$ Eigenvectors (\ref{8}).
This boils down to the equivalence theorem \cite{equiv} between the BFKL and
QCD dipole expression for the 4-gluon amplitude at tree-level.

All in all, the forward one-loop dipole amplitude reads: 
\begin{eqnarray}
f_{Q=0} (\rho,\rho^{\prime};Y)\!\!\!&& = \alpha ^4\ \left(\frac {\alpha N_C}{%
\pi}\right)^2 \int \! dh dh^{\prime}dh_a dh_b \ {\cal G}(h,h_a,h_b) {\cal G}%
^*(h^{\prime},h_a,h_b)  \nonumber \\
&&\left[(\rho^{\prime})^{h^{\prime}-\frac12}(\bar \rho^{\prime})^{\tilde h%
^{\prime}-\frac12} \rho^{-h+\frac12}\bar \rho^{-\tilde h+\frac12}\right]
\int d^2q \left(\frac {\bar q}2\right)^{h^{\prime}-h-1} \left(\frac {q}2%
\right)^{\tilde h^{\prime}- \tilde h-1}  \nonumber \\
&&\int_0^y\! d\bar y \int_0^{Y-y}\! d\bar y^{\prime}\int\! d\omega d\omega_1
d\omega^{\prime}d\omega^{\prime}_1 \frac {e^{\omega \bar y + \omega_1 (y-%
\bar y)}} {(\omega(h_a)\!+\!\omega(h_b)\!-\!\omega)(\omega_1\!-\!\omega(h))}
\nonumber \\
&&\ \ \ \ \ \ \times\ \frac {e^{\omega^{\prime}\bar y^{\prime}+
\omega^{\prime}_1 (y^{\prime}-\bar y^{\prime})}}{(\omega(h_a)\!+\!%
\omega(h_b)\!-\!\omega^{\prime})(\omega^{\prime}_1\!-\!\omega(h^{\prime}))}\
.  \label{34}
\end{eqnarray}
The integration over $d^2q$ yields a $\delta (h-h^{\prime}) \equiv
\delta_{n,n^{\prime}} \delta (\nu-\nu^{\prime}).$ Integrating over $%
h^{\prime}$ and noting that $\mid \!{\cal G}(h,h_a,h_b)\! \mid^2 = \left\vert 
\frac {g_{3{\cal P}} (h,h_a,h_b)} {b(h_a)b(h_b)b(h)} \right\vert^2,$ we
finally get

\begin{eqnarray}
f_{Q=0} (\rho,\rho^{\prime};Y)\!\!\!&& = \alpha ^4\ \left(\frac {\alpha N_C}{%
\pi}\right)^2\int \! dh dh_a dh_b \ \left\vert \frac {g_{3{\cal P}}
(h,h_a,h_b)} {b(h_a)b(h_b)b(h)} \right\vert^2  \nonumber \\
&&\ \ \ \ \ \ \times \left[\left(\frac {\rho^{\prime}}{\rho}\right)^{h-\frac1%
2} \left(\frac {\bar \rho^{\prime}}{\bar \rho}\right)^{\tilde h-\frac12}
\right]  \nonumber \\
&&\int_0^y \!d\bar y \int_0^{Y\!-\!y}\! d\bar y^{\prime}\int\! d\omega
d\omega_1 d\omega^{\prime}d\omega^{\prime}_1 \frac {e^{\omega \bar y +
\omega_1 (y-\bar y)}} {(\omega(h_a)\!+\!\omega(h_b)\!-\!\omega)(\omega_1\!-%
\!\omega(h))}  \nonumber \\
&&\ \ \ \ \ \ \times\frac {e^{\omega^{\prime}\bar y^{\prime}+
\omega^{\prime}_1 (y^{\prime}-\bar y^{\prime})}}{(\omega(h_a)\!+\!%
\omega(h_b)\!-\!\omega^{\prime})(\omega^{\prime}_1\!-\!\omega(h))}\ .
\label{36}
\end{eqnarray}

It is worthwhile to notice that the quantity between brackets in formula (%
\ref{36}) is nothing but $E^h_Q (\rho^{\prime})\ \bar E^h_Q(\rho)$ at $%
Q\equiv 0,$ corresponding to the forward $SL(2,{\cal C})$ Eigenvectors in
the mixed representation (cf. (\ref{8})). The generalization of (\ref{36})
to the non-forward amplitude amounts to replace the quantity between
brackets by the expression for arbitrary $Q.$ This is the global conformal
invariance of the 4-gluon amplitude. We have explicitely checked \cite{nav2}
this property which is thus an hint of the global conformal invariance of
the 4-gluon amplitude maintained at one-loop level in the dipole formalism.

The integration over rapidity variables yields two different contributions
depending on the sign of the quantity $\omega(h_a)\!+\!\omega(h_b)\!-\!%
\omega(h).$ Indeed for $\omega(h_a)\!+\!\omega(h_b)\!<\!\omega(h),$ the
relevant poles are situated at $\omega\!=\!\omega_1\!=\!\omega^{\prime}\!=\!%
\omega^{\prime}_1 =\omega(h).$ leading to expression (\ref{14}) which is
associated with the single Pomeron dependence $e^{\omega(h)\ Y}.$ In the
opposite case, namely $\omega(h_a)\!+\!\omega(h_b)\!>\!\omega(h),$ the
relevant poles are situated at $\omega\!=\!\omega_1\!=\!\omega^{\prime}\!=\!%
\omega^{\prime}_1 =\omega(h_a)\!+\!\omega(h_b).$ The resulting amplitude is
given by (\ref{16}), which corresponds to the double-Pomeron energy
behaviour $e^{(\omega(h_a)\!+\!\omega(h_b))\ Y}.$ Notice that either
expression depends only on the sum $Y=y+y^{\prime},$ as it should from
longitudinal boost invariance.
\bigskip

{\bf 3.} Let us discuss the physical interpretation of the two components (%
\ref{14}) and (\ref{16}) of the one-loop dipole amplitude. At high $Y,$ the
behaviour of these components is driven by the dominant $n=0$ Eigenvalue of
the function $\omega,$ see Eqn. (\ref{6}), for the appropriate argument ($h$
or $h_a,h_b,$ depending on the component). For the first component (\ref{14}%
), the leading behaviour is fixed by the integration over $h$ in the
vicinity of the BFKL Pomeron, namely $n\equiv 0, \nu \approx 0, \omega(h)
\approx \omega (1/2) = \frac {4 \bar \alpha N_c \log 2 }{\pi}.$ the
integration over $h_a,h_b$ remains free provided $\omega (h_{a,b}) < \omega
(1/2).$ A natural interpretation is that it corresponds to the one-loop
dipole contribution $f^{({\cal P})}_Q (\rho,\rho^{\prime};Y)$ to the BFKL
Pomeron.

The dominant $Y-$behaviour of the second component (\ref{16}) is fixed by
integration over $h_a,h_b,$ namely $n_a=n_b=0,$ and $\nu_a,\nu_b \approx 0,$ 
$\omega(h_a)+\omega(h_b) \approx 2\ \omega(1/2) = \frac {8 \bar \alpha N_c
\log 2 }{\pi}.$ This behaviour is typical of a double Pomeron contribution.
The interpretation is the Pomeron-Pomeron cut unitarity correction $f^{(%
{\cal P}\otimes{\cal P})}_Q (\rho,\rho^{\prime};Y)$ following from the
one-loop dipole amplitude. This second component was already estimated \cite
{mu3} in the forward direction. The form of the resulting amplitude was the
same, with an unknown factor which has to be identified now with the triple
Pomeron vertex squared. Another result of the exact calculation is that the
integral over the conformal quantum numers $h$ has to be performed on all
values of the conformal spin $n,$ and has no reason to be restricted to $n=0,
$ as found in \cite{mu3}.

The resulting one-loop dipole contributions to the elastic 4-gluon amplitude
are related to the triple Pomeron vertex function $\left\vert g_{3{\cal P}%
}(h\!,\!h_a\!,\!h_b)\right\vert^2$ which also appears in the perturbative
QCD derivation of hard diffraction \cite{bia}. On a QCD theoretical ground,
the strength of hard diffraction is indeed connected to the sadddle-point
value $g_{3{\cal P}}\left(\frac 12,\frac 12,\frac 12\right)$ which is known
to be a large coefficient ($\approx 7700$, see \cite{triplep}).

Comparing the one-loop results (\ref{14},\ref{16}) with the tree-level
amplitude (\ref{4}), the dipole loop is of order $\bar \alpha ^2$. This is
expected for a Pomeron-Pomeron cut contribution to the elastic amplitude (%
\ref{18}). On the other hand, the dipole loop correction (\ref{14}) to the BFKL Pomeron is thus only of next-next leading order. However it is useful to
notice that the integrand may be quite large and a summation over the
conformal quantum numbers has to be performed. A more quantitative estimate
certainly deserves further study. The presence of the denominator $(\chi
(h)-\chi (h_a)-\chi (h_b))^2$ leads to series of double poles. For the
component $f^{({\cal P})}_Q$ the double poles are solutions of $\chi
(h)\approx 8\log2,$ value which corresponds to the saddle points at $%
h_a=h_b\approx \frac 12.$ The component $f^{({\cal P}\otimes{\cal P})}_Q $
corresponds to summation over both $h_a$ and $h_b$ satisfying the relation $%
\chi (h_a)+\chi (h_b)\approx 4\log 2.$ Note that this double summation may
be related to the toroidal geometry of a one-loop string amplitude which
often appears in string theory calculations as double summations and thus
may give some new insight on the stringy nature of an effective theory \cite
{pe}.

In conclusion, we hope that the analytical calculation of the double BFKL
pomeron exchange in the dipole formulation may open the way to a derivation
of the unitarity corrections to the bare pomeron at weak coupling. Indeed,
it has been remarked \cite{mu1} that the full unitary elastic amplitude
cannot be reconstructed from approximate evaluations of the multi pomeron
exchanges. Analytical solutions of the multi pomeron amplitudes are thus
likely to be evaluated following the same line as the present paper.
Concerning the higher order corrections to the bare pomeron trajectories the
situation is more intricate. Knowing that the next-leading order correction
is large and of order $\bar \alpha $ \cite{next}, the physical meaning and
relevance of the dipole contribution $f^{({\cal P})}_Q$ has to be clarified.
These matters certainly deserve work to be done in the future.

{\bf Acknowledgements}

\noindent We acknowledge fruitful discussions with Al Mueller, Misha Ryskin
and Samuel Wallon and Edmond Iancu for a careful reading of the manuscript.

\eject
{\bf FIGURES}

\bigskip \input epsf \vsize=8.truecm \hsize=10.truecm \epsfxsize=8.cm{%
\centerline{\epsfbox{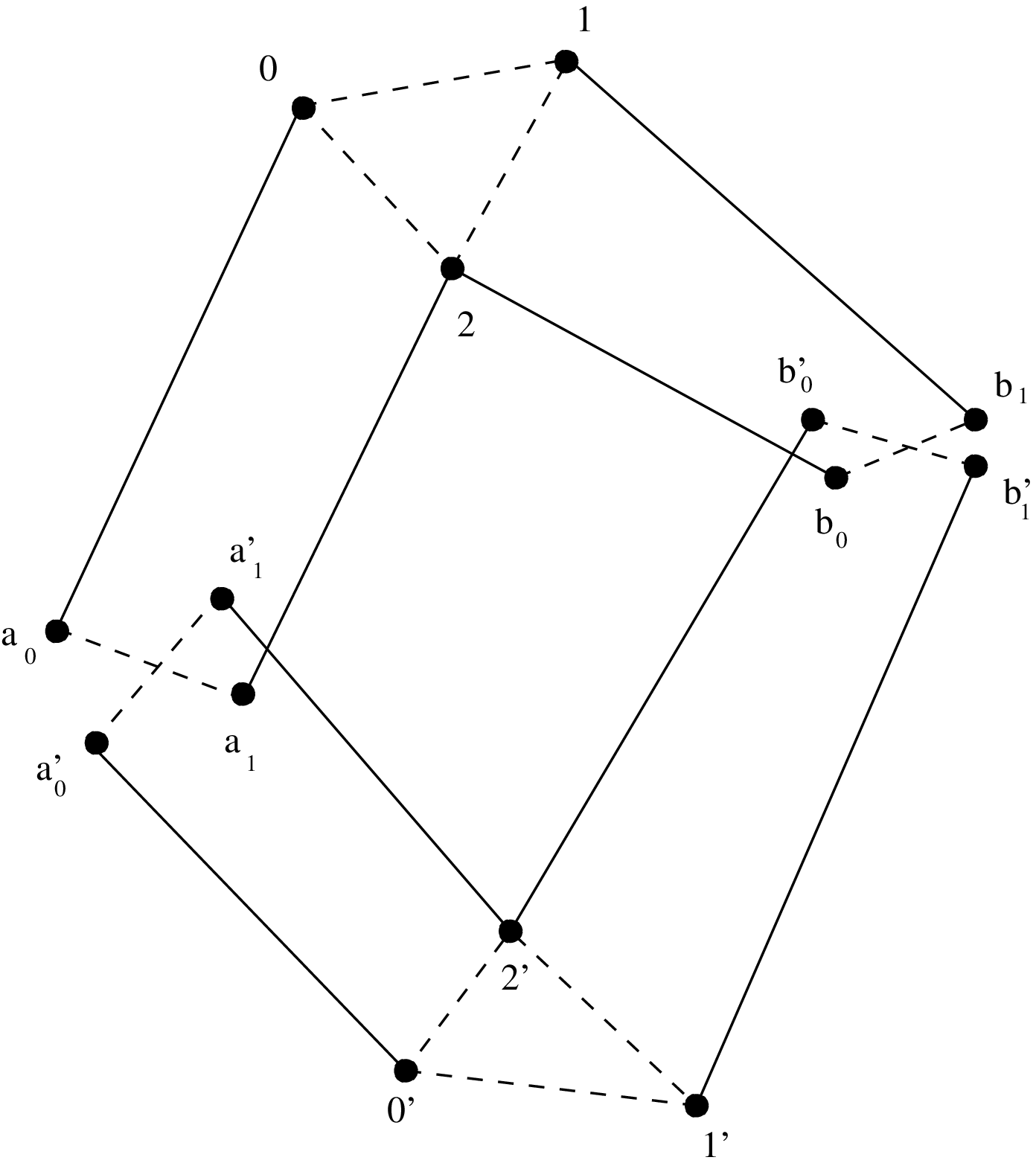}}} {\bf Figure 1-a}

{\it One-loop dipole amplitude in transverse coordinate space}

The transverse coordinates of the incident dipoles ($\rho _{0},\!\rho
_{1},\! \rho^{\prime}_{0},\!\rho^{\prime}_{1}$) are denoted by their indices
together with the two interacting dipoles. The coordinates $\rho _{2},
\rho^{\prime}_{2}$ refer to the points where two independent dipole
evolutions start.
\bigskip  

\input epsf \vsize=8.truecm \hsize=10.truecm \epsfxsize=8.cm{%
\centerline{\epsfbox{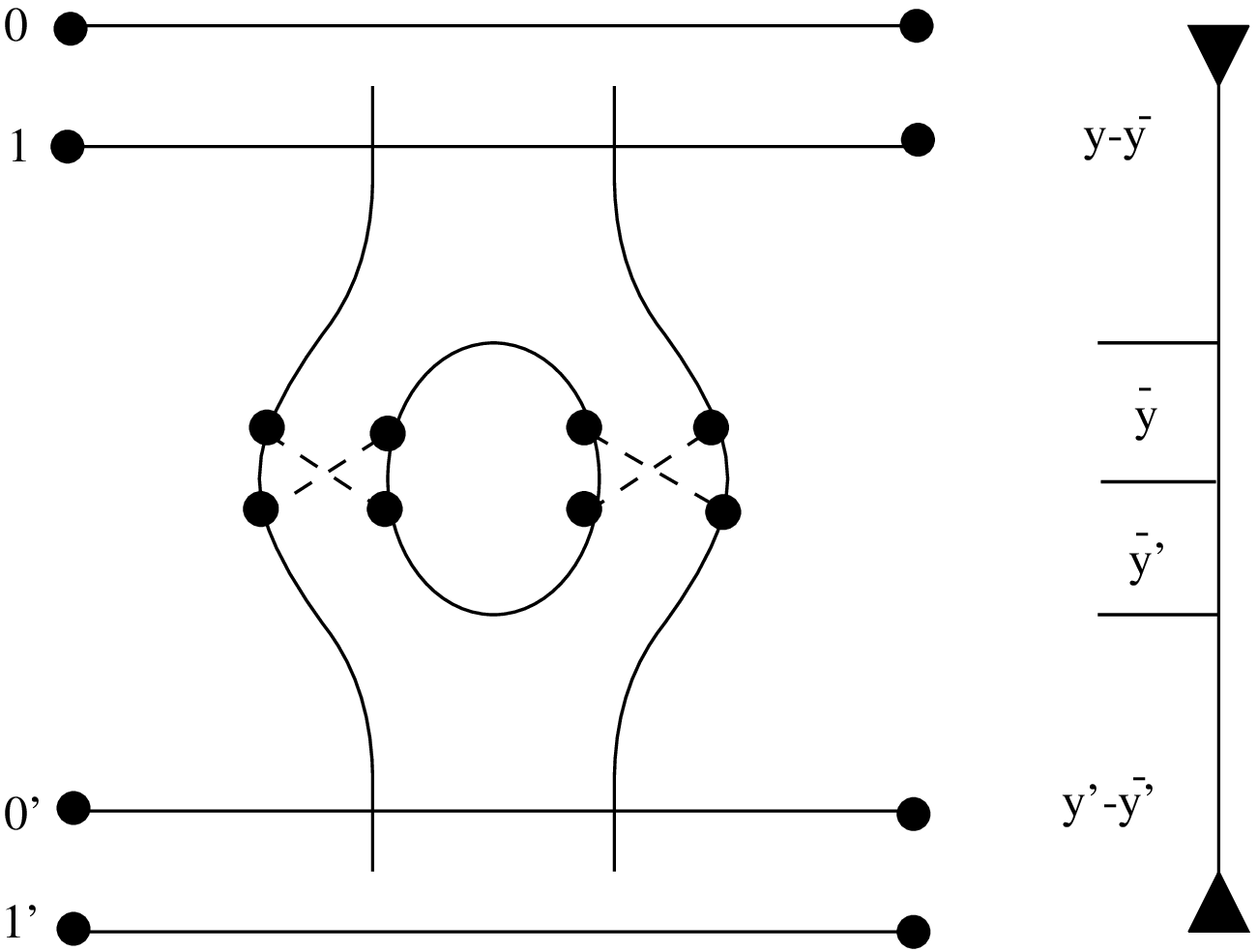}}}\bigskip 
 {\bf Figure 1-b} 

{\it One-loop dipole amplitude in rapidity space}
The one-loop dipole amplitude is sketched in rapidity space. The initial
dipoles $\rho _{0}\rho _{1}, \rho^{\prime}_{0}\rho^{\prime}_{1}$ evolve with
rapidity $y-\bar y^{\prime},y^{\prime}-\bar y^{\prime}$ with one-Pomeron type
of evolution and then with two-Pomeron type of evolution during $\bar y,
 \bar y^{\prime}.$ Note that $y+y^{\prime}\equiv Y,$ the total available
rapidity. After integration over $\bar y, \bar y^{\prime},$ (see text) the
resulting amplitude depends only on $Y$. Note that the amplitude $f^{({\cal P})},$ formula (\ref {14}) corresponds to $\bar y = \bar y^{\prime} \approx 0,$ while the two-Pomeron component $f^{({\cal P}\otimes{\cal P})},$ formula (\ref {16}) corresponds to $\bar y + \bar y^{\prime} \approx Y.$

\end{document}